\documentclass[aps,pra,twocolumn,showpacs,floatfix,superscriptaddress,noeprint]{revtex4-1}
\usepackage[utf8]{inputenc}
\usepackage{color}
\usepackage{amsmath,amssymb,amsfonts,bbm,graphicx,hyperref,times,url}

\usepackage{footnote}

\newcommand{\id}{\mathbbm{1}}

\newcommand{\tr}{{\rm Tr}\,}
\renewcommand{\det}{{\rm Det \,}}
\newcommand{\gr}[1]{\boldsymbol{#1}}
\newcommand{\be}{\begin{equation}}
\newcommand{\ee}{\end{equation}}
\newcommand{\bea}{\begin{eqnarray}}
\newcommand{\eea}{\end{eqnarray}}

\newcommand{\sig}{\gr{\sigma}}

\begin{document}
\title{Locally optimal control of continuous variable entanglement}
\author{Francesco Albarelli}
\affiliation{Quantum Technology Lab, Dipartimento di Fisica {``Aldo Pontremoli''}, Universitá degli Studi di
 Milano, I-20133 Milano, Italy}
\affiliation{Department of Physics and Astronomy, University College London, Gower Street, London WC1E 6BT, UK}
\author{Uther Shackerley-Bennett}
\affiliation{Department of Physics and Astronomy, University College London, Gower Street, London WC1E 6BT, UK}
\author{Alessio Serafini}
\affiliation{Department of Physics and Astronomy, University College London, Gower Street, London WC1E 6BT, UK}

\begin{abstract}
We consider a system of two bosonic modes each subject to the dynamics induced by a 
thermal Markovian environment and we identify instantaneous, local symplectic controls that minimise the loss of entanglement in the Gaussian regime. 
By minimising the decrease of the logarithmic negativity at every instant in time,
it will be shown that a non-trivial, finite amount of local squeezing helps to counter the effect of decoherence during the evolution.
We also determine optimal control routines in the more restrictive scenario where the control operations are applied on only one of the two modes.
We find that applying an instantaneous control only at the beginning of the dynamics, i.e. preparing an appropriate initial state, is the optimal strategy for states with symmetric correlations and when the dynamics is the same on both modes. More generally, even in asymmetric cases, the delayed decay of entanglement resulting from the optimal preparation of the initial state 
with no further action turns out to be always very close to the optimised control where multiple operations are applied during the evolution.
Our study extends directly to `mono-symmetric' systems of any number of modes, i.e. to systems that are invariant under any local permutation of the modes within any one partition, as they are locally equivalent to two-mode systems.
\end{abstract}

\maketitle

\section{Introduction and background: The control of quantum entanglement}

The property of entanglement exhibited by quantum systems has both metaphysical and technological interest. In applications, it has been found to be a fragile resource that is quickly lost when the system is not isolated. Since any manipulation of quantum systems requires them to be in contact with a noisy environment, a major question in advancing control upon them -- arguably the main directive towards the development of 
functional quantum technologies --  is how we can preserve this phenomenon for as long as it takes for an experiment
to unfold.
The design and application of quantum control techniques aimed at sustaining the entanglement of quantum systems has thus been a lively 
area of work over the last fifteen years \cite{raimond01,malinovsky04,mancio07,yamamoto08,serafozzi10,nurdin12,creffield07,carvalho07,carvalho08,carvalho11,thomas05,barbosa10,yan11,masada15,martin17,deng17} 
that has seen the exploration of open-loop \cite{malinovsky04,creffield07} as well as measurement-based \cite{mancio07,yamamoto08,serafozzi10,nurdin12,carvalho07,carvalho08,carvalho11}
and quantum coherent feedback \cite{yan11} strategies, applicable in principle to a wide variety of systems, although quantum optical scenarios seem to offer accurate enough control and low enough noise to facilitate such endeavours \cite{raimond01,thomas05,deng17}. 

In this paper, we address the open-loop control of Gaussian entanglement, such as the one 
displayed in experiments based on parametric down conversion processes \cite{thomas05,barbosa10,deng17}. As well as 
providing an insightful theoretical landscape where, as we shall see, much can be evaluated analytically 
even when realistic noise is included, 
Gaussian systems are widely applicable not only to optical set-ups, but also in all other situations where the interaction 
between constituents and with the environment may be linearised, e.g. in optomechanics, ion traps, atomic ensembles 
and certain quantum circuits \cite{serafozzi17}. 
In particular, we shall consider a two-mode system where each mode is subject to independent loss and thermal noise, as would be the case in degenerate parametric processes that give rise to two-mode squeezed states, the most representative of entangled Gaussian states \cite{thomas05,barbosa10}.
Alongside the free, noisy evolution of an initial entangled state, which clearly degrades the entanglement, we shall consider 
the possibility of acting, at any instant in time, with arbitrary, impulsive 
local symplectic transformations. Let us remind the reader that 
such transformations, that will be assumed to be instantaneous, correspond to all local unitary operations that preserve the Gaussian character of the state. 
This assumption is not unreasonable, if one considers, for instance, loss rates in the range of $10-10^3$ ${\rm kHz}$ compared 
to manipulation times of the order of $1-10$ ${\rm ns}$, both achievable simultaneously in practice in a number of systems. 
This flavour of `optimal' control, whereby some figure of merit was locally optimised over instantaneous unitary manipulations, 
was already applied to discrete quantum systems in \cite{victor13}, was first extended to 
(single-mode) Gaussian states in \cite{carlini14}, and was then adapted to the 
control of the global entropies of multimode Gaussian states in \cite{uther17}. 
Here, we further extend this approach to the non-trivial question of controlling quantum entanglement of a two-mode system.

In this study, we will be able to determine the optimal form of the local symplectic transformation that,
{\em at each given time}, 
minimises the loss in entanglement in terms of the logarithmic negativity, a suitable quantifier.
Quite remarkably, we shall see that a single manipulation of the state through optimal local symplectic transformations 
is always optimal in cases with enough symmetry (either the same loss and thermal noise on both modes or  initial states with symmetric correlations)
: in such cases no further impulsive transformations are required.
It is worth mentioning that the application of a single initial Gaussian control 
to delay the loss of nonclassical properties of non-Gaussian states under the effect of decoherence has been considered previously in the literature ~\cite{Serafini2004a,Serafini2005,Filip2013,Jeannic2017}.
Interestingly, we also have evidence that applying the control repeatedly only gives marginal improvements in the general case of asymmetric dynamics and correlations.
Furthermore, we shall also consider the case where only one of the two modes can be controlled, showing that our analytical conditions extend to this case too.

There is, of course, no guarantee that this
`time-local' optimal control, where the first derivative with respect to time of the figure of merit is optimised at all instants, would achieve a global optimisation.
Hence, we shall test it against a different control strategy, where an alternative quantifier related to the entanglement of the evolving two-mode Gaussian state is optimised at all times. We find that the control based on the logarithmic negativity proves more robust: in our numerical investigation, we could find no way of outperforming it.
We will also quantify the advantage our optimised control 
grants in certain practical cases, in terms of ebits of logarithmic negativity.

The paper is organised as follows. Section~\ref{sec_Gau} introduces the notation and concepts to describe Gaussian systems, with an emphasis on symplectic invariants (that will play 
a prominent role in obtaining our results); this will include a reminder on the characterisation of Gaussian quantum entanglement as well as 
on the diffusive noise model we will consider;
in Section~\ref{sec_Opti}, we derive the locally-optimal control strategy; in Section~\ref{sec_repeated} we discuss the effectiveness of a single initial control operation, while in Section~\ref{sec_single} we turn to the case of single-mode local control; Section~\ref{sec_plots} is devoted to the assessment of the quantitative advantage of our control schemes in concrete cases; in Section \ref{sec_Alte}, a different locally-optimal strategy is considered, and shown to be surpassed by our original scheme; finally, Section \ref{sec_Outro} contains a summary of results and some conclusive remarks.

\section{Gaussian quantum systems and dynamics}\label{sec_Gau}
\subsection{Gaussian states and symplectic invariants}

We introduce here the necessary definitions and tools concerning Gaussian systems, following \cite{serafozzi17}.

Let $\hat{\mathbf{r}} = (\hat{x}_1,\hat{p}_1,\ldots, \hat{x}_n,\hat{p}_n)^{\sf T}$ be a vector of canonical operators such that $[\hat{x}_j,\hat{p}_k] = i\delta_{jk}$, where $\delta_{jk}$ is the Kronecker delta. The canonical commutation relations may also be expressed as 
$[\hat{\mathbf{r}},\hat{\mathbf{r}}^{\sf T}] = i\Omega$, where the commutator is to be interpreted as an outer product between vectors and the anti-symmetric, symplectic form is given by
\begin{equation}
\Omega = \bigoplus_{i=1}^n \begin{pmatrix} 0 & 1 \\ -1 & 0 \end{pmatrix}.
\end{equation}
Second-order Hamiltonians are defined as those that can be written as $\hat{H} = \frac{1}{2}\hat{\mathbf{r}}^{\sf T} H \hat{\mathbf{r}} + \hat{\mathbf{r}}^{\sf T} \mathbf{a}$ where $H$ is a $2n\times 2n$, real, symmetric matrix and $\mathbf{a}$ is a vector of real numbers. The set of Gaussian states can be defined as the ground and thermal states of positive-definite quadratic Hamiltonians:
\begin{equation}\label{eq:system}
\hat{\rho}_G = \frac{e^{-\beta \hat{H}}}{\tr[e^{-\beta \hat{H}}]},
\end{equation}
where $\beta $ is the inverse temperature of the state. Note that the definition above encompasses the limiting instances of the inverse temperature $\beta$, which are needed to describe modes in pure quantum states. Such states are referred to as Gaussian due to their Wigner representation which takes a Gaussian form. As is well known, a Gaussian state $\hat{\rho}_G$ is completely characterised
by the first- and second-order statistical moments of the canonical operators. First moments may be adjusted arbitrarily by local unitary 
`displacement' operations, and thus their values are inconsequential to the entanglement of a quantum state. We will therefore focus on 
the second moments alone, which are usually grouped together in the so-called covariance matrix:
\begin{equation}
\sig = \tr[\{(\hat{\mathbf{r}}-\mathbf{d}),(\hat{\mathbf{r}}-\mathbf{d})^{\sf T}\}\hat{\rho}_G] \; ,
\end{equation}
where $\mathbf{d} = \tr[\hat{\mathbf{r}}\hat{\rho}_G]$ and 
the anticommutator is taken on the outer product of vectors of operators and the trace acts on the Hilbert space, thus obtaining 
a $2n \times 2n$ covariance matrix $\sig$. In order to be a {\em bona fide} covariance matrix, $\sig$ must abide by the 
uncertainty principle, $\sig+i\Omega\ge 0$.

A unitary operation sends Gaussian states into Gaussian states if and only if it is generated by a second-order Hamiltonian.  
Since we are disregarding the first moments, we can set ${\bf a}=0$ in the definition above and thus obtain the group of symplectic 
transformations, generated at the Hilbert space level by purely quadratic Hamiltonians. A symplectic transformation $S$ 
describes the linear, Heisenberg-picture evolution of the vector $\hat{\mathbf{r}}$ under such a purely quadratic Hamiltonian, as 
$\hat{\mathbf{r}}\mapsto S \hat{\mathbf{r}}$. 
Since they correspond to unitary operations, symplectic transformations must preserve the canonical commutation relations: they can in fact be defined as the set of $2n \times 2n$ matrices $S$ such that
\begin{equation}
S\Omega S^{\sf T} = \Omega \; ,
\end{equation}
which form the real symplectic group $\operatorname{Sp}_{2n,\mathbb{R}}$.
Covariance matrices transform under the finite dimensional representation of the symplectic group by conjugation:
\begin{equation}
\sig \mapsto S\sig S^{\sf T}.
\end{equation}
In the study at hand, we will focus on two-mode states and consider control strategies enacted through local symplectic transformations, 
acting separately on each mode, which belong to the direct sum $\operatorname{Sp}_{2,\mathbb{R}}
\oplus\operatorname{Sp}_{2,\mathbb{R}}$. Explicitly, a generic local symplectic transformation $S_\text{loc}$ on two modes  
may be written as 
\begin{equation}\label{symploco}
S_\text{loc} = \left(\begin{array}{cc}
S_1 & 0 \\
0 & S_2 
\end{array}\right) \; , \quad {\rm with} \quad S_1,S_2 \in \operatorname{Sp}_{2,\mathbb{R}} .
\end{equation} 
In the quantum optical practice, such operations correspond to sequences of single-mode squeezers and 
phase shifters (also known as ``phase plates'').

All the spectral properties of Gaussian states are clearly invariant under unitary, and hence symplectic, transformations, 
and must thus be determined by the symplectic invariants of the covariance matrix \cite{dodonov,serafozzi06}. 
As a consequence of the normal mode decomposition, 
that allows one to turn any covariance matrix $\sig$ into normal form through some symplectic transformation $S$, as per 
$S \sig S^{\sf T} = \bigoplus_{j=1}^{n} \nu_j {\mathbbm 1}$, $n$ such 
independent invariants can be constructed from an $n$-mode covariance matrix. 
A possible choice of symplectic invariants is represented by the $n$ quantities $\nu_j$, known as the symplectic eigenvalues, 
that may be determined as the moduli of the eigenvalues of the matrix $\Omega\sig$ which, due to the symmetry of 
$\sig$ and anti-symmetry of $\Omega$, come in degenerate pairs. In terms of the symplectic eigenvalues, the uncertainty relation may be expressed as $\nu_j\ge 1$.

The information contained in the symplectic eigenvalues may be expressed in terms of other sets of invariants, such as 
the sums of the $2k\times 2k$ principal minors of the matrix $\Omega\sig$, which we shall denote with $\Delta_{k}^n$. Up to a sign, these correspond to the coefficients of the characteristic polynomial of $\Omega\sig$. 
It turns out that the uncertainty relation implies the inequality \cite{serafozzi06}
\be
\Sigma = \sum_{k=0}^{n} (-1)^{n+k} \Delta_{k}^n \ge 0 \; ,
\ee
to be read with the additional stipulation $\Delta_0^n=1$.
For two-mode states, the independent invariants are the determinant ${\rm Det}\sig=\Delta_2^2$ and the 
quantity $\Delta_1^2={\rm Det }\boldsymbol{\alpha}+{\rm Det}\boldsymbol{\beta}+2{\rm Det}\boldsymbol{\gamma}$ \cite{serafozzi04}, for the $2\times 2$ submatrices 
$\boldsymbol{\alpha}$, $\boldsymbol{\beta}$ and $\boldsymbol{\gamma}$ defined by 
\be
\sig = \begin{pmatrix}
\boldsymbol{\alpha} & \boldsymbol{\gamma} \\
\boldsymbol{\gamma}^\mathsf{T} & \boldsymbol{\beta}
\end{pmatrix} \; .
\label{eq:sigME_blocks}
\ee

The properties related to the entanglement of Gaussian states are instead determined by quantities that are invariant under 
{\em local} symplectic transformations. Since the positivity of the partial transposition is necessary and sufficient for the separability 
of $1$ vs $n$ and locally symmetric Gaussian states, it turns out that a satisfactory characterisation of the entanglement of such states,
which encompass all two-mode states, 
is achieved in terms of partially transposed symplectic eigenvalues and invariants. At the level of covariance matrices, 
the partial transposition $\tilde{\sig}$ of the first $n_A$ modes 
is described by the congruence transformation $\tilde{\sig}=T \sig T$ for $T=(\bigoplus_{j=1}^{n_A} \sigma_z) \oplus \id_{2(n-n_A)}$,
where $\sigma_z$ is the Pauli matrix ${\rm diag}(1,-1)$ and $\id_{2(n-n_A)}$ is the identity matrix on the remaining $(n-n_A)$ modes.
For this set of Gaussian states, the violation of the inequality $\tilde{\sig}+i\Omega$ is necessary and sufficient for entanglement. 
In terms of the sums of principal minors of $\Omega\tilde{\sig}$, denoted with $\tilde{\Delta}_{k}^n$, one has the corresponding 
necessary and sufficient separability condition
\be
\tilde{\Sigma} = \sum_{k=0}^{n} (-1)^{n+k} \tilde{\Delta}_{k}^n \ge 0 \; . \label{invasepa}
\ee
An equivalent necessary and sufficient condition for separability may be written down in terms of the smallest partially transposed symplectic eigenvalue of $\sig$, which we shall denote with $\tilde{\nu}_{-}$:
\be
\tilde{\nu}_{-} \ge 1 \; .
\ee
For the class of states in hand, it turns out that at most one partially transposed symplectic eigenvalue may be smaller than $1$.
In turn, $\tilde{\nu}_{-}$ determines the logarithmic negativity $E_{{\mathcal N}}$ of the state, as
\be
E_{{\mathcal N}} = \max\{0,-\log_2(\tilde{\nu}_{-})\} \; . \label{logneg}
\ee
The quantity $E_{{\mathcal N}}$ sets an upper bound, expressed in entangled bits (ebits), to the distillable entanglement,
i.e.\ the rate of maximally entangled pairs of qubits that can be distilled through local operations and classical communication 
from the Gaussian state in the asymptotic limit of an infinite number of available copies \cite{vidalwerner,plenio05,myungshik}. 
This is the entanglement quantifier 
that we shall adopt as a figure of merit for our control scheme.

For two-mode states, the partially transposed invariants are the determinant ${\rm Det}\sig$ which, by virtue of Binet's theorem, 
is not affected by the partial transposition $T$, and $\tilde{\Delta}_1^2 ={\rm Det \alpha}+{\rm Det}\beta - 2{\rm Det}\gamma$; notice the minus sign that distinguishes this quantity from $\Delta_1^2$. This yields the separability criterion 
\be
\tilde{\Sigma} = {\rm Det}\sig - \tilde{\Delta}_1^2 +1 \ge 0 \, , \label{invasepa2m}
\ee
and the smallest partially transposed symplectic eigenvalue $\tilde{\nu}_-$ determined by the relationship
\be
2\tilde{\nu}_{-}^2 = \tilde{\Delta}_1^2 - \sqrt{(\tilde{\Delta}_1^2)^2-4{\rm Det}\sig} \; . \label{noumeno}
\ee

\subsection{Open system dynamics with thermal environments}\label{subsec_Dyn}
As the free, uncontrolled, evolution of the system, responsible for the loss of coherence and quantum entanglement, 
we shall adopt the rotating wave interaction with a number of bosonic Markovian environments 
with a different coupling strength and number of thermal excitations for each mode, 
resulting in the following Lindblad master equation
\begin{equation}
\dot{\hat{\rho}} = \sum_i^N \chi_i \mathcal{D}\left[ \hat{a}_i \right] \! \hat{\rho} + \left( \chi_i - \gamma_i \right) \mathcal{D} \left[ \hat{a}_i^\dag \right] \! \hat{\rho}
\end{equation}
where the superoperator $\mathcal{D} \left[ \hat{o} \right] \! \hat{\rho} = \hat{o} \hat{\rho} \hat{o}^\dag - \left\{ \hat{o}^\dag \hat{o} , \hat{\rho} \right\} $ was introduced; this dynamics is often referred to as `loss in a thermal environment', in this case a different thermal environment for each mode. 
The parameters $\gamma_i$ are the loss rates (actually, we could omit one by scaling the unit of time), whilst $\chi_i= \gamma_i ( \exp[\frac{\hbar\omega_	i}{k T_i} ]+1)/(\exp[\frac{\hbar\omega_	i}{k T_i} ]-1) = \gamma_i \left( 2 N_i + 1 \right)$, in terms of the baths' temperatures $T_i$, 
the modes frequencies $\omega_i$, Planck's constant $\hbar$ and Boltzmann's constant $k$; alternatively in terms of the mean number of bosonic excitations of the baths $N_i =1/(\exp[\frac{\hbar\omega_	i}{k T_i} ]-1)$.
We remark that the number of excitations $N_i$ (and thus the parameters $\chi_i$) are different for bosonic modes at different frequencies even if the temperature is the same. 
This is relevant in various physical setups, e.g. in opto-mechanical systems.

In terms of the covariance matrix, this dynamics leads to a diffusive equation:
\begin{equation}\label{eq:covmastergen}
\dot{\sig} = A \sig + \sig A^\mathsf{T} + D \; ,
\end{equation}
with $A = -\frac{1}{2}\bigoplus_i^n  \gamma_i \id_2$ and $D = \bigoplus_i^n  \chi_i \id_2 $. 
Throughout the paper, the notation $\id_{j}$ will indicate the $j\times j$ identity matrix.
We remark that Eq.~\eqref{eq:covmastergen} holds true for any state; to completely determine the evolution of a Gaussian state one also needs the equation for first moments: $\dot{\mathbf{d}}= A \mathbf{d}$.
As previously explained, all entanglement properties of Gaussian states are encoded in the covariance matrix.

Given the block form \eqref{eq:sigME_blocks}, we explicitly write the differential equation~\eqref{eq:covmastergen} for a two-mode state:
\bea \label{eq:dot_sig_blocks}
& \dot{\sig} = \begin{pmatrix}
\dot{\boldsymbol{\alpha}} & \dot{\boldsymbol{\gamma}} \\
\dot{\boldsymbol{\gamma}}^\mathsf{T} & \dot{\boldsymbol{\beta}}
\end{pmatrix}  = 
&\begin{pmatrix}
- \gamma_1 \boldsymbol{\alpha} + \chi_1 \id_2  & - \frac{\gamma_1 +\gamma_2}{2}  \boldsymbol{\gamma}\\
- \frac{\gamma_1 +\gamma_2}{2}  \boldsymbol{\gamma}^\mathsf{T} & - \gamma_2 \boldsymbol{\beta} + \chi_2  \id_2
\end{pmatrix}  ,
\eea
moreover we can easily write the solution $\sig(t)$ in terms of the submatrices
\begin{align}
\boldsymbol{\alpha}(t) = & \,\frac{\chi_1}{\gamma_1} \id_2 + \left( \boldsymbol{\alpha}(0) - \frac{\chi_1}{\gamma_1} \id_2 \right) e^{- \gamma_1 t} \label{eq:evol_alpha} \\ 
\boldsymbol{\gamma}(t) = & \, \boldsymbol{\gamma}(0) e^{- \frac{\gamma_1 +\gamma_2}{2} t} \label{eq:evol_gamma} \\
\boldsymbol{\beta}(t) = & \, \frac{\chi_2}{\gamma_2} \id_2 + \left( \boldsymbol{\beta}(0) - \frac{\chi_2}{\gamma_2} \id_2 \right) e^{- \gamma_2 t} \; .\label{eq:evol_beta}
\end{align}

A crucial step for our analysis is to find the equations governing the evolution of the symplectic invariants under this dynamics. 
This can be achieved thanks to Jacobi's formula: $\frac{d}{dt} \det A(t) = \tr \! \left[ \operatorname{Adj}(A(t)) \, \frac{d}{dt} A(t)\right ]$, where $\operatorname{Adj}(A(t))$ represents the adjugate matrix of $A(t)$; for invertible matrices we have $\operatorname{Adj}(A(t)) = \left( \mathrm{Det}A(t) \right) A(t)^{-1}$.
For two modes the equations for the two invariants of the partially transposed state $\tilde{\Delta}_2^2={\rm Det}\sig$ and 
$\tilde{\Delta}_1^2$ can be explicitly written as 
\begin{align}
\dot{{\rm Det}\sig} =& \left( \det \sig \right) \, \tr \! \left[ \sig^{-1} \dot{\sig} \right] =  \label{detdot} \\
=& - 2 \left( \gamma_1 + \gamma_2 \right) \det \sig + \nonumber \\
& + \chi_1 \det \boldsymbol{\beta} \;  \mathrm{Tr}\left[  \sig / \boldsymbol{\beta} \right] + \chi_2 \det \boldsymbol{\alpha} \; \mathrm{Tr} \left[ \sig / \boldsymbol{\alpha} \right] \nonumber \\ 
 \dot{\tilde{\Delta}}_1^2 =& \frac{d}{d t }\bigl( \det \boldsymbol{\alpha} + \det \boldsymbol{\beta} - 2 \det \boldsymbol{\gamma} \bigr)  = \label{deltadot} \\ 
 = & -2 \gamma_1 \det \boldsymbol{\alpha} - 2 \gamma_2 \det \boldsymbol{\beta} + 2 \left( \gamma_1 + \gamma_2 \right) \det \boldsymbol{\gamma} \, + \nonumber \\
& + \chi_1 \tr \boldsymbol{\alpha} + \chi_2 \tr \boldsymbol{\beta} \; , \nonumber
\end{align}
where we used the time derivative of the submatrices~\eqref{eq:dot_sig_blocks}, formulas for the inverse of a block matrix and the fact that $\mathrm{Tr} \left[ \mathrm{Adj} \left( A \right) \right] = \tr A $ for $2 \times 2$ matrices
\footnote{To obtain equation~\eqref{detdot} we also assumed $\det \boldsymbol{\gamma} \neq 0$, because a multiplication by $\boldsymbol{\gamma}^{-1}$ is needed, this is always possible since $\det \boldsymbol{\gamma} < 0$ is a necessary condition to have a nonseparable state.}.
The two matrices $\sig / \boldsymbol{\beta} = \boldsymbol{\alpha} - \boldsymbol{\gamma} \boldsymbol{\beta}^{-1} \boldsymbol{\gamma}^\mathsf{T} $ and $ \sig / \boldsymbol{\alpha} = \boldsymbol{\beta} - \boldsymbol{\gamma}^\mathsf{T} \boldsymbol{\alpha}^{-1} \boldsymbol{\gamma} $ are the Schur complements of $\sig$ with respect to the submatrices $\boldsymbol{\beta}$ and $\boldsymbol{\alpha}$.


The relationships (\ref{detdot},\ref{deltadot}) form the basis of our analysis for the control of entanglement. Notice that, on the right hand side, whilst all the determinants are local symplectic invariants and are thus unaffected by local symplectic transformations, all the traces that multiply the coefficients $\chi_1$ and $\chi_2$ are not. These terms identify the handle through which local 
symplectic control may act.

\section{Optimal time-local control}\label{sec_Opti}

We will now assume the possibility of performing instantaneous, local symplectic transformations, of the form given by 
Eq.~(\ref{symploco}), which cannot generate quantum correlations but can, as we shall see, delay their decay during 
the interaction with the thermal bath described by the diffusive dynamics (\ref{eq:covmastergen}).  

To start with, observe that, since we are assuming arbitrary local symplectic control, we can restrict to covariance matrices 
in Simon normal form, which can be reached through local symplectic transformations \cite{simon00,serafozzi17}, and read:
\be
\sig_\mathsf{nf} = \left(\begin{array}{cccc}
a&0&c_+&0 \\
0&a&0&c_- \\
c_+&0&b&0 \\
0&c_-&0&b 
\end{array}\right) \; .
\ee
Here, $a$, $b$, $c_+$ and $c_-$ may be taken as the four independent local symplectic invariants, 
in terms of which the optimisation problem we set to solve may be cast. Such invariants may be promptly related to a set of quantities, such as 
${\rm Det}\sig$, ${\rm Det}\boldsymbol{\alpha}$, ${\rm Det}\boldsymbol{\beta}$ and ${\rm Det}\boldsymbol{\gamma}$, whose invariance is 
manifest since any local symplectic matrix $S$ has ${\rm Det}S=1$. Since local rotations are symplectic transformations that preserve the local covariance matrices, we may swap $c_+$ and $c_-$ and so without loss of generality assume $c_+\ge c_-$.

As we saw above, the logarithmic negativity of a two-mode Gaussian state is determined by the smallest 
partially transposed symplectic eigenvalue $\tilde{\nu}_-$, which is in turn given by Eq.~(\ref{noumeno})
in terms of the quantities ${\rm Det}\sig$ and $\tilde{\Delta}_1^2={\rm Det}\sig_1+{\rm Det}\sig_2-2{\rm Det}\sig_{12}= a^2+b^2-2c_+c_-$. Our control objective is to minimise the derivative of $2\tilde{\nu}_{-}^2$ as the logarithmic negativity $E_{\mathcal N}$ is a decreasing function of it. It is thus useful to work out explicitly its derivatives 
with respect to ${\rm Det }\sig$ and $\tilde{\Delta}_1^2$:
\begin{align}
\partial_{{\rm Det}\sig} (2\tilde{\nu}_{-}^2) &=  \frac{2}{\sqrt{(\tilde{\Delta}_1^2)^2-4{\rm Det}\sig}} = \frac{2}{u} \, \label{detpar} , \\
\partial_{\tilde{\Delta}_1^2} (2\tilde{\nu}_{-}^2) &=  1 - \frac{\tilde{\Delta}_1^2}{\sqrt{(\tilde{\Delta}_1^2)^2-4{\rm Det}\sig}}  \nonumber \\
&= 1-\frac{a^2+b^2-2c_+c_-}{u} \, , \label{deltapar}
\end{align}
where $u=\sqrt{(a^2-b^2)^2+4ab(c_+^2+c_-^2)-4c_+c_-(a^2+b^2)}$.
Notice also that $u=\sqrt{(\tilde{\Delta}_1^2)^2-4{\rm Det}\sig}$ is always \textit{strictly} greater than zero for entangled states. To see this, 
it will suffice to notice that only one partially transposed symplectic eigenvalue of a two-mode covariance matrix may ever be smaller than $1$ \cite{serafozzi17,serafozzi06}, as is the case for an entangled state. Therefore, the two symplectic eigenvalues of a two-mode entangled state must be different, which is equivalent to stating $u>0$.  

Now, the free, diffusive evolution of the quantities ${\rm Det }\sig$ and $\tilde{\Delta}_1^2$ has been determined above 
and is given by
\begin{align}
\dot{{\rm Det}\sig} &= - 2 \left( \gamma_1 + \gamma_2 \right) \det \sig + \label{detdot_loc} \\
& + \chi_1 \det   \boldsymbol{\beta} \; \mathrm{Tr}\left[ S_1 ( \sig / \boldsymbol{\beta} ) S_1^\mathsf{T} \right] + \chi_2  \det   \boldsymbol{\alpha} \;  \mathrm{Tr} \left[  S_2  ( \sig / \boldsymbol{\alpha} ) S_2^\mathsf{T} \right] \nonumber  \; , \nonumber \\
\dot{\tilde{\Delta}}_1^2 & = -2 \gamma_1 \det \boldsymbol{\alpha} - 2 \gamma_2 \det \boldsymbol{\beta} + 2 \left( \gamma_1 + \gamma_2 \right) \det \boldsymbol{\gamma} \, + \nonumber \\
& \quad + \chi_1 \mathrm{Tr} \left[ S_1 \boldsymbol{\alpha} S_1^\mathsf{T} \right] + \chi_2 \mathrm{Tr} \left[ S_2 \boldsymbol{\beta} S_2^\mathsf{T} \right] \; , \label{deltadot_loc}
\end{align}
where the role of the control has been made explicit through the local symplectic transformations $S_1$ and $S_2$.
We intend to determine the symplectic matrices $S_1$ and $S_2$ that minimise the 
derivative of $\tilde{\nu}_-$ and thus optimally preserve the entanglement over an infinitesimal time-interval; whereby 
our control is termed ``local'' or ``time-local''.


Each of the single-mode symplectic matrices $S_1$ and $S_2$ admits a  
singular value decomposition as $S_j= Q_j Z_j R_j$, where $Q_j$ and $R_{j}$ are orthogonal symplectic matrices and 
$Z_j=\mathrm{diag}(z_j,1/z_j)$ is a local squeezing operation \cite{serafozzi17,uther16}. This fact simplifies our optimisation considerably: first, notice that 
all the terms depending on $S_j$ in Eqs.~(\ref{detdot_loc},\ref{deltadot_loc}) are invariant under local orthogonal transformations, so that the transformations $Q_j$ can be ignored altogether.
Besides, due to our use of the Simon normal form it may be shown that the minimisation we are considering below is always realised 
for $R_j=\id_2$ and a proper choice of $Z_j$. Hence, the only 
relevant local action may come from the local squeezing operations $Z_1$ and $Z_2$, diagonal in the 
normal quadratures (as we shall refer to the degrees of freedom that attain the Simon normal form).

The straightforward evaluation of the quantities depending on the local symplectic transformations in Eqs.~(\ref{detdot_loc},\ref{deltadot_loc}) yields
\begin{align}
\det   \boldsymbol{\beta} \; \mathrm{Tr}\left[ S_1 ( \sig / \boldsymbol{\beta} ) S_1^\mathsf{T} \right] &= \left[ \frac{b (a b - c_-^2)}{z_1^2} + b (a b - c_+^2) z_1^2 \right] \label{m3z} \\
\det   \boldsymbol{\alpha} \;  \mathrm{Tr} \left[  S_2  ( \sig / \boldsymbol{\alpha} ) S_2^\mathsf{T} \right]  &= \left[ \frac{a (a b - c_-^2)}{z_2^2} +  a (a b - c_+^2) z_2^2 \right]  \\
\mathrm{Tr}\left[ Z_1 \boldsymbol{\alpha} Z_1^\mathsf{T} \right] &= \;\; a \left(z_1^2 + \frac{1}{z_1^2}\right)
\label{tralpha} \;\; \\
\mathrm{Tr}\left[ Z_2 \boldsymbol{\beta} Z_2^\mathsf{T} \right ] & = \;\; b \left(z_2^2 + \frac{1}{z_2^2}\right) \;\; .\label{trbeta}
\end{align}

Now, our optimisation over the squeezing parameters $z_1$ and $z_2$ may be defined by considering 
only the additive part in the derivative of $(2\tilde{\nu}_-)^2$ that contain such parameters; 
it will also be convenient to multiply such a term by the positive quantity $u>0$ (that does not depend on $z_1$ 
and $z_2$) and to refer to the quantity obtained by $\xi$. Eqs.~(\ref{detpar},\ref{deltapar}) and (\ref{m3z}-\ref{trbeta}) yield
\be
\label{eq:xi}
\xi =  \chi_1 \left( v_1z_1^2 +\frac{w_1}{z_1^2} \right) + \chi_2 \left( v_2z_2^2 +\frac{w_2}{z_2^2} \right) \, ,
\ee
with
\begin{align}
v_1 &= a (b^2-a^2+u) + 2c_+ (ac_--bc_+) \; , \label{v1} \\
w_1 &= a (b^2-a^2+u) + 2c_- (ac_+-bc_-) \; ,\label{w1}\\
v_2 &= b (a^2-b^2+u) + 2c_+ (bc_--ac_+) \; ,\label{v2}\\
w_2 &= b (a^2-b^2+u) + 2c_- (bc_+-ac_-) \; . \label{w2}
\end{align}
All of the quantities $v_j$'s and $w_j$'s are bound to be positive semidefinite: if that were not the case, 
there would exist physical cases where the free dynamics, that corresponds to local completely positive maps, 
would be able to increase the logarithmic negativity, 
which is impossible since the latter is an entanglement monotone \cite{plenio05}.
In point of fact, it will be expedient to just consider the quantities 
$v_j$ and $w_j$ as strictly positive in what follows. 
The possibility for such parameters to take null values would imply the existence of limiting states where the 
free, thermal lossy channel does not decrease the entanglement: we are not aware of any such states (nor did we 
encounter any numerical evidence where any of the parameters above were zero).

Hence, the minimisation of $\xi$, which yields the optimal local squeezing parameters to 
preserve as much quantum entanglement as possible (in the form of logarithmic negativity), is performed by separately minimising the two terms in~\eqref{eq:xi}, as they are both strictly positive; the dependence on the the two parameters $\chi_j$ thus vanishes.
The result is straightforward,
and furnishes the optimal parameters:
\be
\bar{z}_j = \sqrt[4]{\frac{w_j}{v_j}} \; . \label{zopt}
\ee


\subsection{Initial state preparation versus repeated control}\label{sec_repeated}
A few preliminary considerations concerning the optimal control strategy emerge 
directly from Eqs.~(\ref{v1}--\ref{zopt}). If the magnitudes of the $x$- and $p$-correlations in the normal form 
$c_+$ and $c_-$ are the same, i.e. if $c_+=-c_-$ (note that $c_+$ and $c_-$ of different signs are a prerequisite for entanglement \cite{serafozzi17}), then $v_j=w_j$ and  
$\bar{z}_j=1$ for $j=1,2$.
The optimal control strategy consists therefore just in putting the initial 
state in normal form through local symplectic transformations, and then in letting the state evolve freely,
since the evolution we are considering will keep the covariance matrix in normal form.

However, if $|c_+|\neq|c_-|$, then the optimal control strategy includes some local squeezing after the normal form reduction; the form of such a locally-squeezed state is not preserved by the dynamics.
Nonetheless, when the two modes undergo the same dynamics, i.e. $\gamma_1 = \gamma_2$ and $\chi_1 = \chi_2$ it turns out that a single iteration of the initial control operation leads to time-local optimal control over the whole dynamics.
To write this in formulae we consider a given optimal covariance matrix ${\sig}'$, i.e. $\sig' = S_{\text{loc}} \sig_0 S_{\text{loc}}^\intercal$ such that $S_{\text{loc}}$ minimises $\dot{\tilde{\nu}}(S_{loc} \sig_0 S_{loc}^{\sf T})$ and $\sig_0$ is a completely generic initial physical covariance matrix. The statement above is then equivalent to
\be
\inf_{S_{loc}} \dot{\tilde{\nu}}\left(S_{loc}\left(p \sig'+(1-p)\chi\id_4\right)S_{loc}^{\sf T}\right) = 
\dot{\tilde{\nu}}\left(p \sig'+(1-p)\chi\id_4 \right) ,
\ee
$\forall$ $p\in [0,1]$, with $S_{loc}$ optimised over the set $\operatorname{Sp}_{2,{\mathbbm R}}\oplus \operatorname{Sp}_{2,{\mathbbm R}}$.
The convex combination appearing in the previous equation is the diffusive time evolution of Eqs.~(\ref{eq:evol_alpha}-\ref{eq:evol_beta}) with the choice $\gamma_i= \gamma$, $\chi_i = \chi$, a rescaling of time $t \to \gamma t$ and upon the substitution $p={\rm e}^{- t}$. 
The statement above is proven by showing that the local squeezing of the matrix $p\sig'+(1-p)\chi\id_4$ matches,
for all $p$, the optimal squeezing condition \eqref{zopt}, where the quantities (\ref{v1}-\ref{w2}) must be re-evaluated 
for $p\sig'+(1-p)\chi\id_4$.
This yields and equality in terms of the quantities $a$, $b$, $c_+$ and $c_-$ of the initial state, 
whose explicit algebraic expression is utterly unwieldy, but which can be verified exactly with Mathematica.

In general, when $\chi_1 \neq \chi_2 $ and $\gamma_1 \neq \gamma_2$ the previous statement does not hold.
Therefore we can quantitatively evaluate the effectiveness of repeated control operations applied during the considered dynamics.
Numerical evidence (presented in Section~\ref{sec_plots}) shows that subsequent control operations after the initial one improve the conservation of entanglement, but by a very small amount.
The intuitive reason behind for this behaviour is that the repeated scheme gets more useful the greater the unbalance between the values $|c_+|$ and $|c_-|$.
However, an unbalance between those correlation terms is achieved only for mixed states and, due to the uncertainty principle, that restricts physical covariance matrices, there is only a narrow region in parameter space where such unbalanced states contain substantial entanglement~\cite{adesso04}.

\begin{figure}[t!]
\centering
\includegraphics[width=\columnwidth]{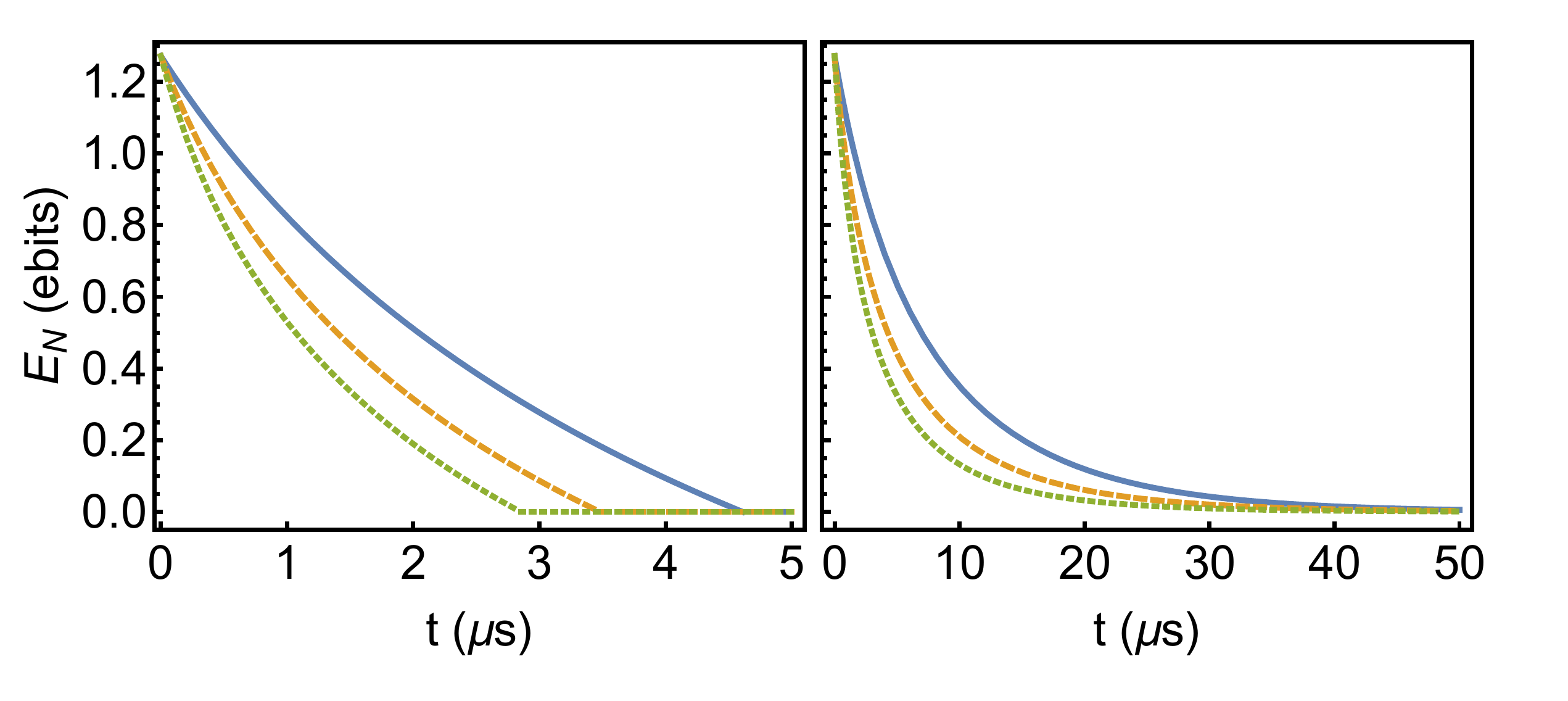}
\caption{Logarithmic negativity of states which are initially locally equivalent to a pure two-mode squeezed state with squeezing parameter $r=\ln[\sqrt{2}+1]/2$, evolving in a thermal environment with $\chi_j=2$ (left panel) and $\chi = 1.000013 $, i.e. room temperature at $450$ ${\rm THz}$ (right panel); the two modes have the same loss rate $\gamma = 100 {\rm kHz}$. 
The blue (solid) curve, refers to an initial state in standard form, the orange (dashed) curve refers to an initial state where one mode is locally squeezed 
by the squeezing transformation $\bar{Z}={\rm diag}(2,1/2)$, while the green (dotted) curve refers to a state where both modes 
have been squeezed by $\bar{Z}$. \label{fig:sym_equal}}
\end{figure}

\subsection{Time-local control of a single subsystem}
\label{sec_single}

We now consider a different scenario where we have the ability to control only one of the two subsystems; without loss of generality we assume this to be the first subsystem, the one corresponding to the local covariance matrix $\boldsymbol{\alpha}$.
At variance with the previous case we cannot put the state in normal form, since local symplectic transformations on both parties would be needed.
We can however apply symplectic transformations on one system to make the local covariance matrix $\boldsymbol{\alpha}$ proportional to the identity, and then further act on such a subsystem with a local rotation (which will leave $\boldsymbol{\alpha}$ unaffected but can act on the correlation sub-block). In this case, the free parameters are eight, and it is convenient to parametrize the whole CM in terms of the coefficients of the normal form as follows
\begin{equation}\label{eq:sig_local_alpha}
\sig = S_{\alpha, \beta} \sig_\mathsf{nf} S_{\alpha, \beta}^\mathsf{T} \qquad  S_{\alpha, \beta} = \id_2 \oplus S_\beta
\end{equation}
$S_\beta$ is a generic symplectic transform on the second subsystem.
We can readily see that the ten free real parameters of a generic two-mode covariance matrix are reduced to eight, since $S_\beta$ will depend on two angles and a (unbounded) squeezing parameter.

We can then reproduce all the previous steps, in particular we now have to evaluate Eqs.~(\ref{detdot_loc},\ref{deltadot_loc}) considering only the local symplectic control $S_1 = Z_1 R_1$; i.e. setting $S_2 = \id_2 $.
The minimization proceeds in the same way, with the minimum always attained for $R_1 = \id_2$ and for a diagonal squeezing operation $Z_1$
with squeezing parameter $\bar{z}_1$ given by Eq.~(\ref{zopt}) in terms of the coefficients of the normal form.
Even if the state cannot be put in normal form, the only parameters that come into play are the normal form coefficients, since the local symplectic $S_\beta$ on the initial state does not affect the result. This is due to the fact that, despite the presence of entanglement, 
the contributions of the individual modes to the derivative in Eqs.~(\ref{detdot_loc},\ref{deltadot_loc}) are positive and independent.
Note that the symplectic transformations $S_1$ is the control operation, whilst $S_\beta$ above was only used to parametrise the initial state.



\begin{figure}[t!]
\includegraphics[width=\columnwidth]{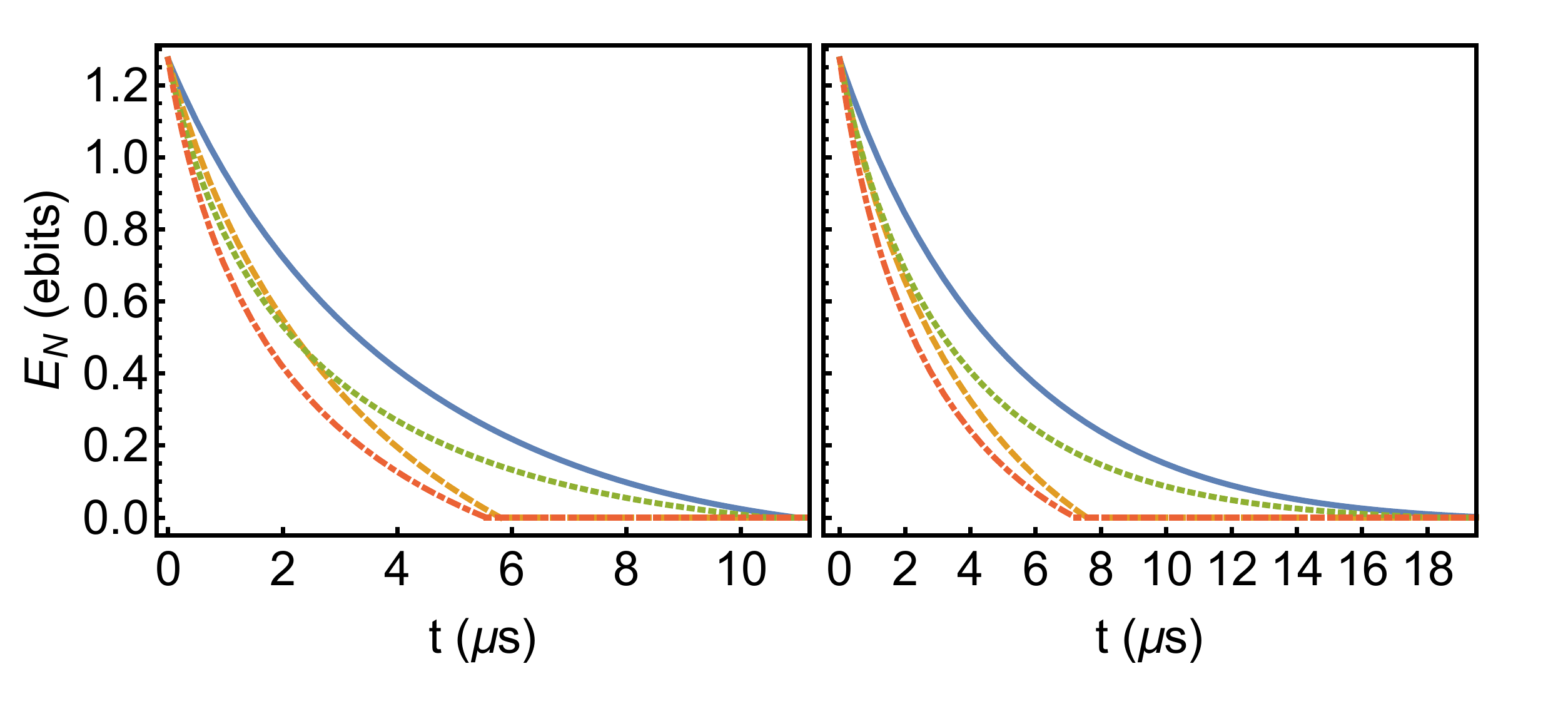}
\caption{Logarithmic negativity of states which are initially locally equivalent to a pure two-mode squeezed state with squeezing parameter $r=\ln[\sqrt{2}+1]/2$.
\emph{Left panel}: the two modes are evolving in two thermal environment with $\chi_1 = 1$ and $\chi_2=2$, while the loss rate $\gamma=100$ ${\rm kHz}$ is the same for both modes. 
\emph{Right panel}: the two modes are evolving in the same thermal environment with $\chi = 1.000013$ but with loss rates $\gamma_1=100 {\rm kHz}$ and $\gamma_2=10 {\rm kHz}$.
The blue (thick) curve, refers to an initial state in standard form, the orange (dashed) curve refers to an initial state where the first mode is locally squeezed by the squeezing transformation $\bar{Z}={\rm diag}(2,1/2)$, the green (dotted) curve refers to a state where the second mode is squeezed by $\bar{Z}$, while the red (dot-dashed) curve refers to a state where both modes have been squeezed by $\bar{Z}$. \label{fig:sym_diff}}
\end{figure}

\subsection{Quantitative examples}
\label{sec_plots}
Let us first consider the paradigmatic example of an initial two-mode squeezed state, with $a=b=\cosh(2r)$ and $c_+=-c_-=\sinh(2r)$, $r$ being the so called two-mode squeezing parameter. 
Such states are the output of degenerate parametric down conversion processes, and possibly 
the most iconic among entangled Gaussian states.
It is therefore interesting to consider their case 
quantitatively, through an explicit example:
the advantage granted by the initial adjustment that sets the state in normal form is 
illustrated in Fig.~\ref{fig:sym_equal}, where equal loss rates of and equal thermal noises are assumed for both modes,
and the application of single-mode squeezing on initial covariance matrices in normal form corresponding 
to a two-mode squeezed state such that $\cosh(2r)=\sqrt{2}$ is considered.
We consider two cases: a single-mode squeezing operation ${\rm diag}(2,1/2)$ applied on only one mode or applied on both modes.
The first case can also be thought as a single-mode control scenario, as described in Section~\ref{sec_single}.
In the case of a high thermal noise $\chi=2$ (left panel of Fig.~\ref{fig:sym_equal}) it is apparent that the initial control may prolong the life of quantum entanglement by around $2$ $\mu{\rm s}$ (this will of course depend on the strength of the single-mode squeezing operation considered).
Reducing $\chi$ to $1.000013$, which corresponds to the room temperature thermal noise 
affecting a mode of visible radiation at $450$ ${\rm THz}$, results in the comparison in the right panel of Fig.~\ref{fig:sym_equal}: as is well known, 
at such low noise the complete demise of entanglement is less sharp, but adjusting the state in normal form still
allows one to gain about $0.22$ ebits of logarithmic negativity after $10$ $\mu{\rm s}$: a very significant improvement.

Fig.~\ref{fig:sym_diff} shows the effect of different decoherence channels on the same initial two-mode squeezed vacuum.
In the left panel we keep the same loss rate $\gamma = 100 \mathrm{kHz}$ but we change the thermal noise affecting the two modes ($\chi_1 = 2$ and $\chi_2 = 1$), while in the right panel we keep the same room temperature thermal noise $\chi = 1.000013$ but we consider two different loss rates $\gamma_1 = 100 \mathrm{kHz}$ and $\gamma_2 = 10 \mathrm{kHz}$.
In this case it becomes evident how the symmetry between the two modes is lost and depending on which one of the two modes is controlled we have different behaviours.  

\begin{figure}[t!]
\includegraphics[width=.87\columnwidth]{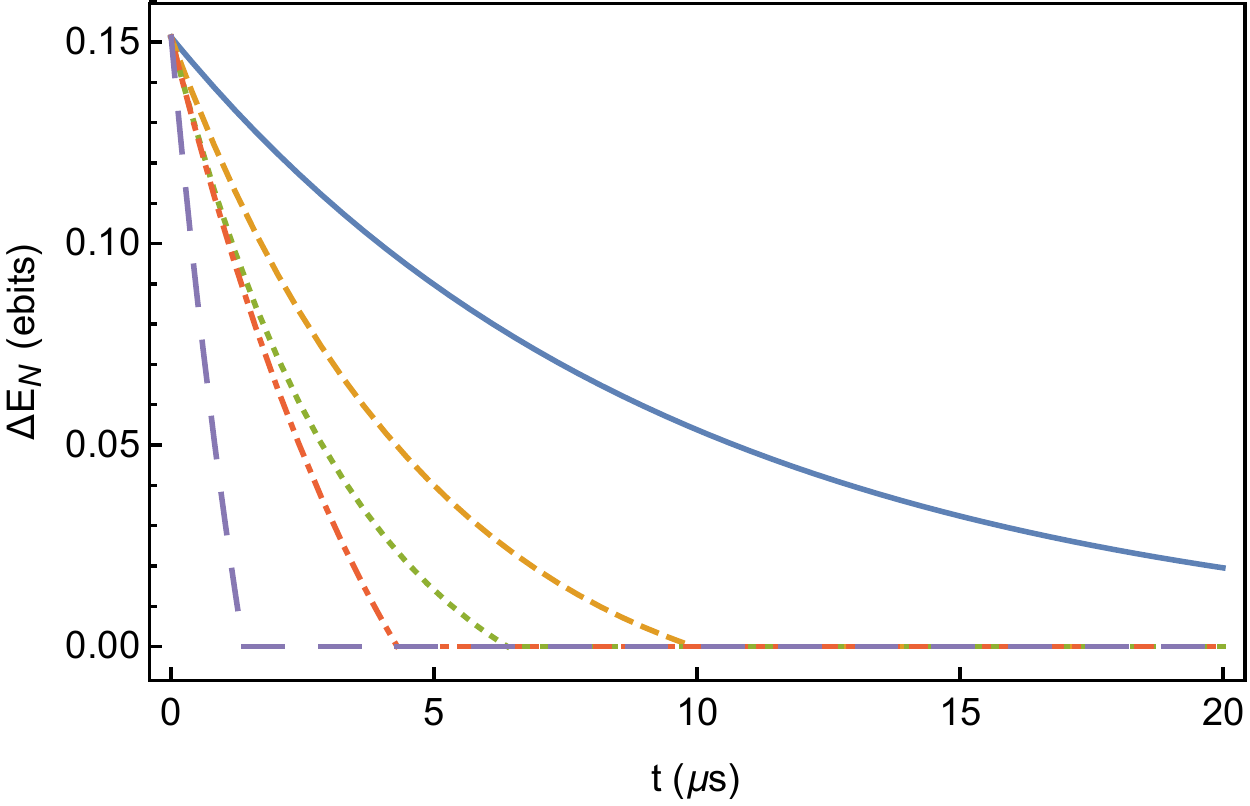}
\caption{Logarithmic negativity for initial states locally equivalent to a state with normal form parameters $a=4.5$, $b=3.5$, $c_+=2.2$, $c_-=-3.5$, evolving in a thermal environment with $\chi=1.000013$ and loss rate of $100$ ${\rm kHz}$. 
The blue (solid) curve pertains to a state that was optimally adjusted on both modes, the orange (short dashed) curve pertains to a state that was optimally adjusted on one mode, the green (dotted) curve pertains to an initial state in normal form, the red (dot-dashed) curve pertains to an initial state whose normal form was altered by 
the squeezing transformation $\bar{Z}={\rm diag}(2,1/2)$ on one mode, 
while the purple (large dashed) curve pertains to a state further altered locally by a phase plate $R_{\frac{\pi}{4}}$ (besides the local squeezing $\bar{Z}$). \label{Asym}}
\end{figure}

Fig.~\ref{Asym} illustrates the effect of a single control step in instances relevant to optical systems, and highlights the influence local operations might have on the dynamics of entanglement under decoherence. In the case depicted, which adopts, as above 
a loss rate of $100$ ${\rm kHz}$ and $\chi=1.000013$ (room temperature at $450$ ${\rm THz}$), the so called 
``sudden death'' of quantum entanglement is delayed from $7$ $\mu{\rm s}$ to a stunning $90$ $\mu{\rm s}$ 
though the optimal control squeezing operation identified above.
It is worth stressing again that the optimal local squeezing transformation does not depend on the parameters of the noise, but only on the 
state in hand.

\subsubsection*{Repeated controls}
As explained previously, repeatedly applying the control during the dynamics is only relevant for initial states with unbalanced correlations $|c_+| \neq |c_-|$ and if the dynamics of the two modes is not the same.
For this quantitative study, we therefore choose thermal environments corresponding to different microwave frequencies of $\omega_1 = 35.0476 \, \mathrm{GHz}$ and $\omega_2 = 55.3674 \, \mathrm{GHz}$, taken from~\cite{Flurin2012}, at room temperature, with equal same loss rates set at $\gamma = 100 \, \mathrm{kHz}$.

\begin{figure}[t!]
\includegraphics[width=.87\columnwidth]{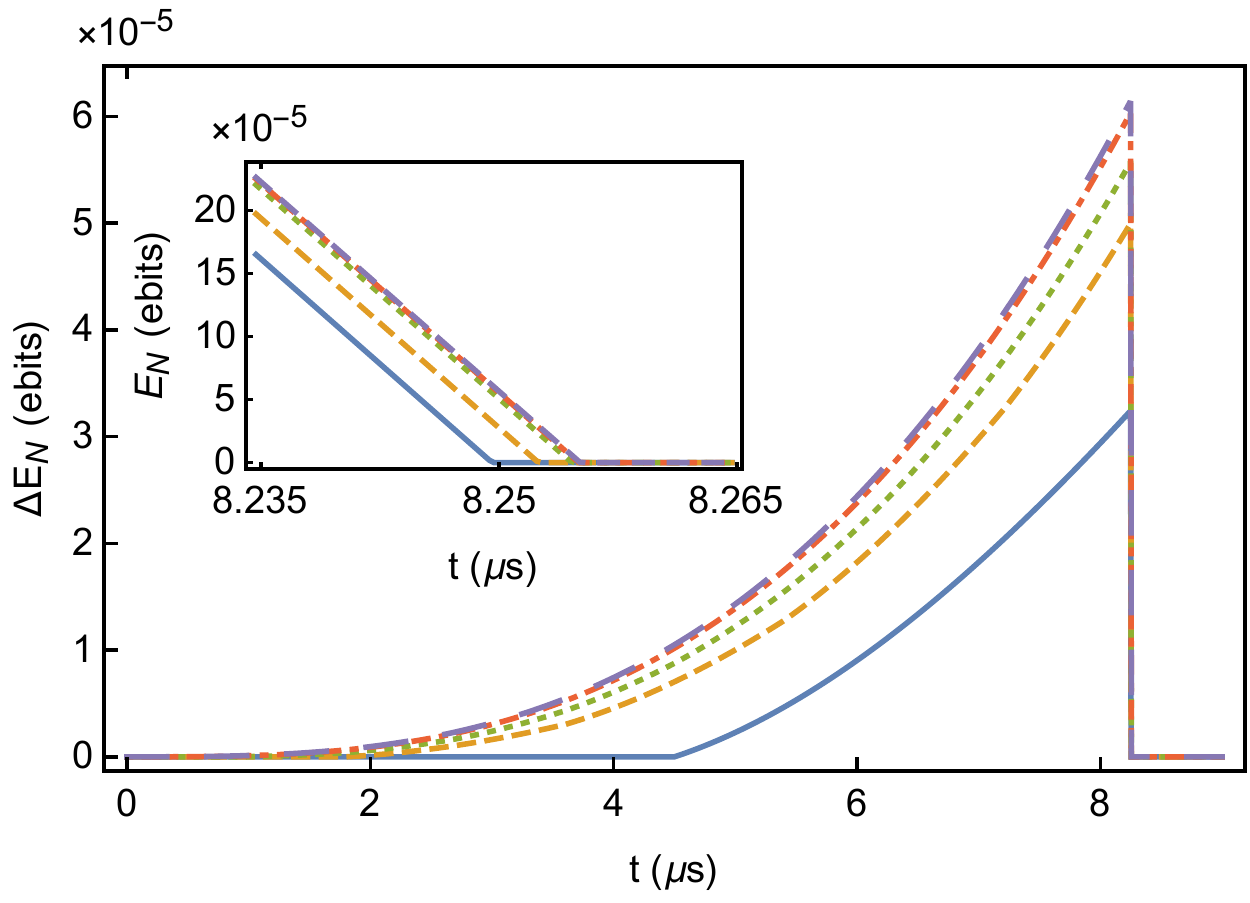}
\caption{Difference between the logarithmic negativity with repeated controls and the one with a single initial control, for an initial state in normal form with parameters $a=4.5$, $b=3.5$, $c_+=2.2$, $c_-=-3.5$, evolving in thermal environments with $\chi_1 / \gamma =1.14769$, $\chi_2 / \gamma = 1.02956$ and same loss rate $\gamma = 100 \, \mathrm{kHz}$.
The blue (solid) curve represent one additional control operations after the first one, the orange (short dashed) curve three additional control operations, the green (dotted) curve represents a total of ten control operations, the red (dot-dashed) curve represents a total of one hundred control operations, while the purple (large dashed) curve is obtained when the control is applied at every time step.
\emph{Inset:} the slight delay of entanglement ``sudden death'' due to repeated controls.
The timestep used in the plots is $2 \times 10^{-3} \mu \mathrm{s}$ and $3 \times 10^{-4} \mu \mathrm{s}$ for the main plot and the inset respectively.
\label{fig:repeated}}
\end{figure}

In Figure~\ref{fig:repeated} we plot the difference between the logarithmic negativity obtained with repeated control operations and the one obtained with a single initial control.
We show that applying more control operations after the initial one delays the quantitative decay of entanglement of a small but not altogether negligible amount; interestingly, the sudden death of entanglement is only very slightly delayed by applying further control operations.
These results are only a particular example of the same general behaviour we found from similar numerical analyses.
However, we also found the effect of repeated control operation is more marked when it is acting on only one mode.

\begin{figure}[t!]
\includegraphics[width=.87\columnwidth]{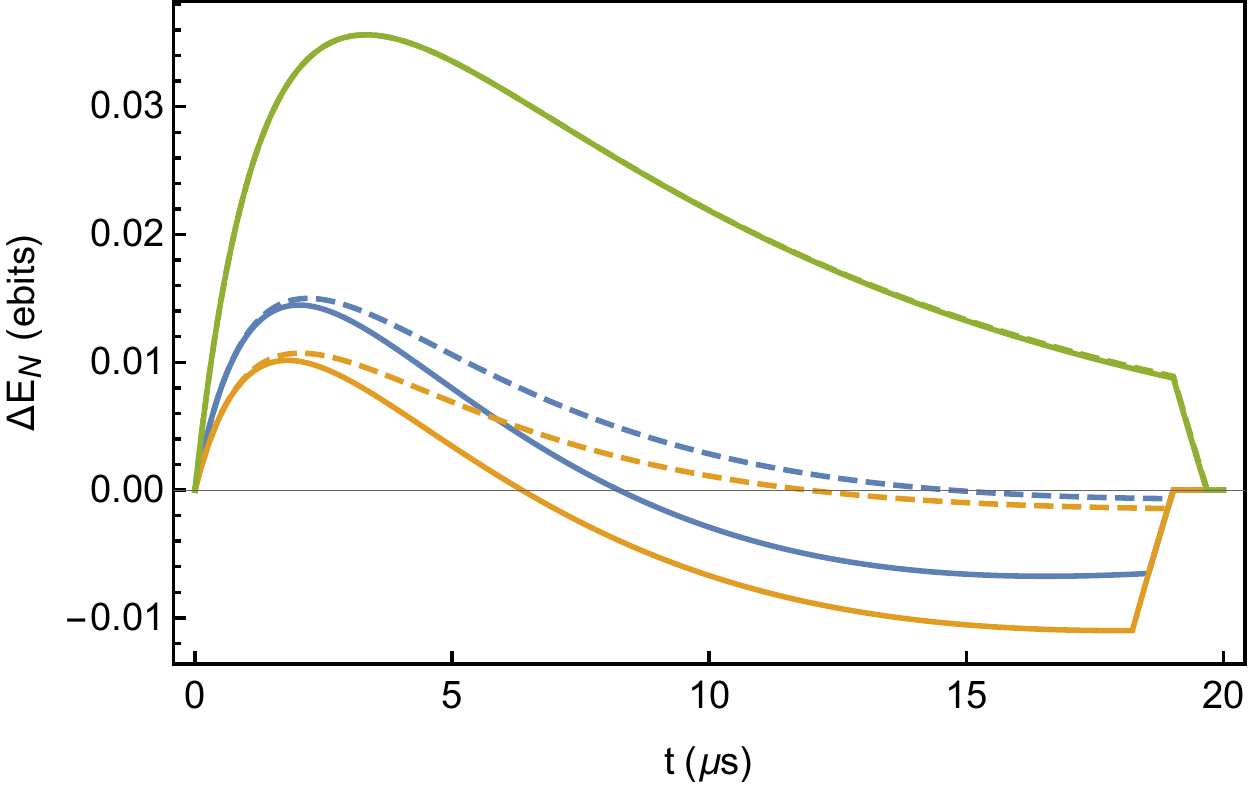}
\caption{Difference between the logarithmic negativity with repeated controls and the one without control, for an initial state with normal form parameters $a=5$, $b=6$, $c_+=5.2$, $c_-=-4.8$, evolving in two thermal environment with $\chi_1 / \gamma =1.14769$, $\chi_2 / \gamma = 1.02956$ and same loss rate $\gamma = 100 \, \mathrm{kHz}$. 
The green curves above all the others represent control operations on both modes, the blue curves in the middle (in the $\approx 2 \, \mathrm{\mu s}$ region ) represent a control operation on subsystem 1, while the orange curves at the bottom (again, in the $\approx 2 \, \mathrm{\mu s}$ region) pertain to control operations on subsystem 2 only.
For every case the full curve represents an initial control operation only, while the dashed curve represents controls applied at each time of the evolution (the timestep chosen for the plot is $10^{-2} \mu \mathrm{s}$).
\label{fig:repeated2}}
\end{figure}

To illustrate that interspersed, time-local optimal control actions might fail to optimise globally, let us report on a peculiar case, 
with normal form parameters $a=5$, $b=6$, $c_+=5.2$ and $c_-=-4.8$,
in Figure~\ref{fig:repeated2}, where we plot the difference between the logarithmic negativity obtained with repeated control operations and the one where no controls are applied and the initial state in normal form remains in normal form throughout the evolution.
We see that applying the control on both modes always provides one with an increased logarithmic negativity and that, as shown before, controlling more than once gives little improvements. 
On the other hand, optimised control operations on a {\em single} mode show that, whilst initially offering a certain advantage in terms of logarithmic negativity,  
can actually make it decrease sooner than without any control operations!
It is also apparent that in this case repeated control operations have a more marked effect.  

\subsection{Alternative local optimisation}\label{sec_Alte}

As previously said the control procedures we identified in the previous sections do not necessarily achieve a global optimisation of logarithmic negativity.
It is therefore interesting to explore alternative locally-optimal controls and compare them to what we achieved above.
As one such alternative, we shall consider the time-local minimisation of the quantity 
$\tilde{\Sigma} = {\rm Det}\sig - \tilde{\Delta}_1^2 +1$ whose negativity is a necessary and sufficient signature of entanglement,
as per Eq.~(\ref{invasepa}). 

\begin{figure}[t!]
\includegraphics[width=.87\columnwidth]{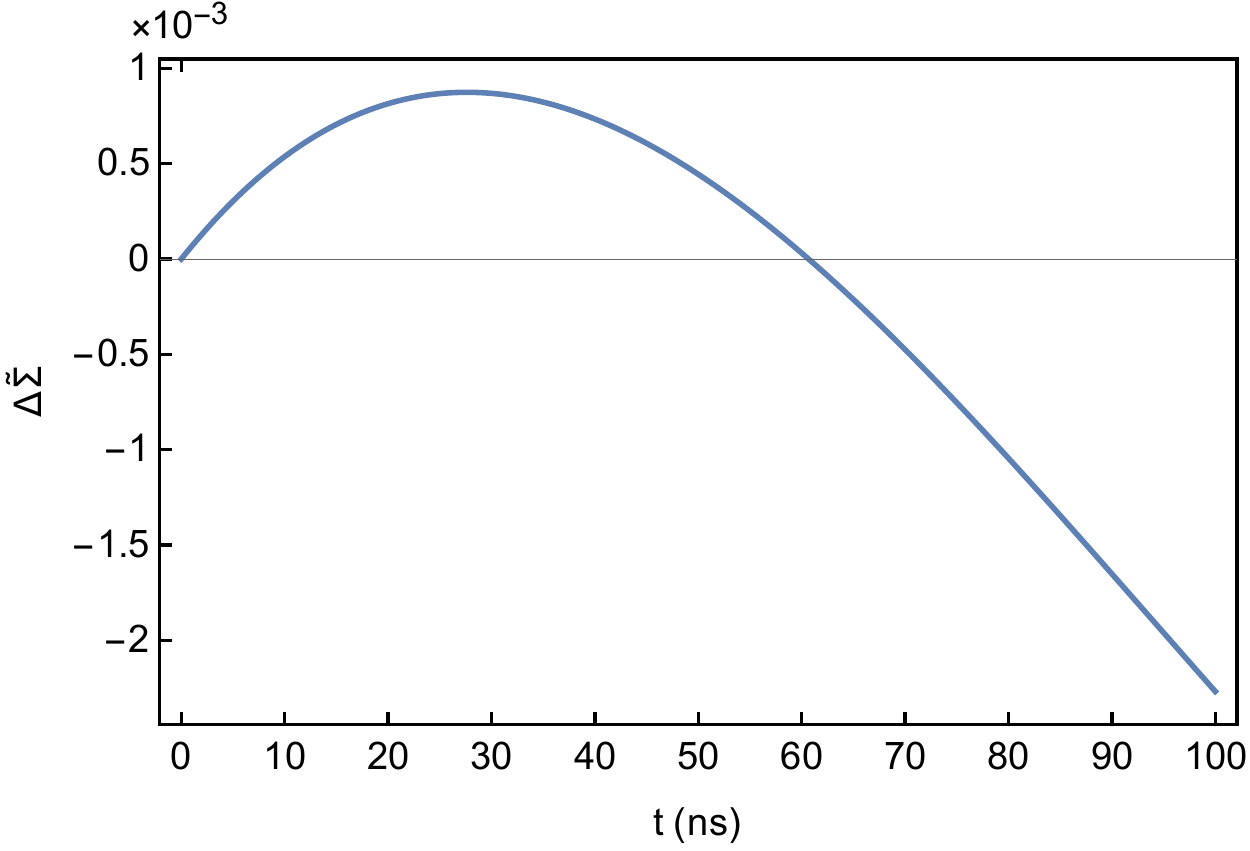}
\caption{Difference in the separability parameter $\tilde{\Sigma}$ between optimal local control minimising
$\tilde{\nu}_{-}$, 
and optimal local control minimising $\tilde{\Sigma}$
for an initial state in normal form with $a=4.5$, $b=3.5$, $c_{+}=2.2$, 
$c_{-}=-3.5$ evolving in an environment with equal thermal noises $\chi_{i}=2$ and loss rates $\gamma_{i} = 1 {\rm MHz}$.
As apparent, after about $30$ ${\rm ns}$, the local minimisation of $\tilde{\nu}_{-}$ offers an advantage also in minimising 
$\tilde{\Sigma}$, whose time-local minimisation is thus shown to be globally suboptimal. 
The evolving state is entangled up to a time of around $90$ ${\rm ns}$.}\label{Alterna} 
\end{figure}

One simply has $\dot{\tilde{\Sigma}} = \dot{\det \sig } - \dot{\tilde{\Delta}}_1^2$, where the two terms are the ones found previously in Eqs.~(\ref{detdot},\ref{deltadot}).  
This leads to a quantity depending on the control parameters $z_1$ and $z_2$,
which will be termed $\xi' = \chi_1 \left( \det   \boldsymbol{\beta} \; \mathrm{Tr}\left[ S_1 ( \sig / \boldsymbol{\beta} ) S_1^\mathsf{T} \right] - \mathrm{Tr} \left[ S_1 \boldsymbol{\alpha} S_1^\mathsf{T} \right] \right) + \chi_2  \left( \det   \boldsymbol{\alpha} \;  \mathrm{Tr} \left[  S_2  ( \sig / \boldsymbol{\alpha} ) S_2^\mathsf{T} \right] 
 -  \mathrm{Tr} \left[ S_2 \boldsymbol{\beta} S_2^\mathsf{T} \right] \right) $.
Explicitly, we are interested in the minimisation of
\be
\xi' =  \chi_1 \left( v_1' z_1^2 +\frac{w'_1}{z_1^2} \right) + \chi_2 \left( v'_2z_2^2 +\frac{w'_2}{z_2^2} \right) \, ,
\ee
with
\begin{align}
v'_1 &= b(ab-c_+^2)-a \; , \label{v1b} \\
w'_1 &= b(ab-c_-^2)-a \; ,\label{w1b}\\
v'_2 &= a(ab-c_+^2)-b \; ,\label{v2b}\\
w'_2 &= a(ab-c_-^2)-b \; , \label{w2b}
\end{align}
which is achieved for the choice 
\be
z_j = \sqrt[4]{\frac{w'_j}{v'_j}} \; . \label{zoptdash}
\ee
This optimisation, at variance with the one detailed in the previous section, 
requires a non-trivial control on-the-fly to be enforced at every instant in time, for every choice of initial states and dynamics (even symmetric ones).

However, rather interestingly, a direct numerical comparison reveals that this control criterion is not globally optimal in order to minimise $\tilde{\Sigma}$ since, as shown in a specific case in Fig.~\ref{Alterna}, we found it is beaten by the control that minimises $\tilde{\nu}_{-}$ locally: in general, the control minimising $\tilde{\nu}_{-}$ is always superior in delaying the disappearance of 
entanglement in a thermal environment, although the difference 
between the two methods is marginal in the region around disentanglement \footnote{That one of such two inequivalent methods should be suboptimal globally, 
could be inferred, barring the case that they would lead to the very same disentanglement time, 
by the fact that both the condition $\tilde{\Sigma}<0$ and $\tilde{\nu}_{-}<1$ are 
necessary and sufficient for entanglement.}. 
In the case depicted in Fig.~\ref{Alterna} of an initial state in normal form with $a=4.5$, $b=3.5$, $c_{+}=2.2$, $c_{-}=-3.5$ evolving in an environment with thermal noise $\chi=2$ and loss rate of $1$ ${\rm MHz}$,
the disappearance of entanglement is delayed by the optimal control method by only $4$ ${\rm ns}$ with respect to 
the method minimising $\tilde{\Sigma}$ ($9486$ ${\rm ns}$ against $9482$ ${\rm ns}$). Note that the 
difference involved is well detected by our numerical precision.

We have thus encountered an instance where the time-local control of a parameter is not globally optimal. 
The time-local minimisation of 
$\tilde{\nu}_{-}$, offering the added advantage of requiring a single initial manipulation, is certainly to be preferred
to a control minimising $\tilde{\Sigma}$. 
In fact, a broad numerical exploration could identify no procedure that would surpass the time-local minimisation of $\tilde{\nu}$ in maximising the final entanglement, although we do not possess an analytical proof of such optimality.

\section{Summary and outlook}\label{sec_Outro}

Summarising, we have identified a local manipulation through unitary operations, comprised of the  
local transformations that puts the state in standard form followed by local squeezing transformations 
with specific parameters, that guarantees that a two-mode Gaussian state evolving in two generic, and generally different, thermal environments (one for each mode) will experience the minimal possible loss of logarithmic negativity at 
each moment in time.
Very remarkably, for states with symmetric correlations and for the same dynamics on both modes, the optimal scheme only involves an initial local unitary adjustment of the state, with no subsequent action.
Interestingly, even when the initial preparation alone is not optimal, 
we find that the results are quantitatively very close to those obtained with continuous, on-the-fly control. 
Such a strategy has been also compared with alternative control methods (optimising different figures of merits) and found to be superior.
It is worth mentioning that the optimal transformation to be applied does not depend at all on the parameters of the noisy evolution.
This advance in the understanding of the manipulation and control of continuous variable entanglement, besides its inherent theoretical relevance, would be applicable to systems of direct and immediate experimental interest~\cite{thomas05,barbosa10}.
Let us also remark that the analysis we carried out straightforwardly extends to all bipartite multimode Gaussian states which are invariant under permutation of the modes within any one partition, since such states are locally equivalent to two-mode states~\cite{lami16}; notably, this includes all $1$ vs $n$-mode systems.

In the future, this study could be extended to consider a different set of instantaneous operations consisting of passive unitaries but without the restriction of being local, i.e. to the optimisation of entanglement through repeated beam splitters (we can already see from Eqs.~(\ref{detdot},\ref{deltadot}) that phase shifters would not be not useful for the task).
The interest of such a problem lies in the fact that in quantum optics passive optical elements are generally considered to be freely available resources~\cite{tan17,yadin18}, while squeezing operations, either single-mode or two-mode, are the expensive ingredient needed to create entangled states.

Yet another worthwhile extension of the present work would consist in adding an interaction Hamiltonian between the two modes during the dynamics. Since the interaction has the potential to create entanglement, it would be interesting to study its interplay with local controls, possibly on only one of the modes.
This setting will be particularly relevant to optomechanical systems, where the interaction Hamiltonian plays a crucial role in, e.g., sideband driving, and the thermal noise on the mechanical mode tends to suppress nonclassical features.

\bibliography{library}

\end{document}